\documentclass[epj]{svjour}
\usepackage{graphicx}

\begin{document}
%
%%%%%%%%%%%%%%%%%%%%%%%%%%%%%%%%%%%%%%%%%%%%%%%%%%%%%%%%%%%%%%%%%%%%%%%%%%%%%%%%%%%
\title{Electronically--implemented coupled logistic maps}
%

%\titlerunning{Electronically--implemented coupled logistic maps}

\author{Alexandre L'Her \inst{1}
        \and
        Pablo Amil  \inst{1}
        \and 
        Nicolás Rubido  \inst{1} \inst{2}
        \and
        Arturo C. Marti \inst{1}
        \and 
        Cecilia Cabeza \inst{1}
}

%\thankstext{e5}{e-mail: cecilia@fisica.edu.uy}


% \institute{First address \label{addr1}
%            \and
%            Second address \label{addr2}
%            \and
%            \emph{Present Address:} if needed\label{addr3}
% }

\institute{Universidad de la Rep\'ublica, Facultad de Ciencias, Igu{\'a} 4225, Montevideo, Uruguay  \and University of 
Aberdeen, King's College, Institute for Complex Systems and Mathematical Biology, Aberdeen, AB24 3UE, United Kingdom.}

% \author{Pablo Amil}
% \affiliation{Universidad de la Rep\'ublica, Facultad de Ciencias, Igu{\'a} 4225, Montevideo, Uruguay.}
% 
% \author{Nicolás Rubido}
% \affiliation{Universidad de la Rep\'ublica, Facultad de Ciencias, Igu{\'a} 4225, Montevideo, Uruguay.}
% \affiliation{University of Aberdeen, King's College, Institute for Complex Systems and Mathematical Biology, Aberdeen, AB24 3UE, United Kingdom.}
% 
% \author{Arturo C. Mart\'{\i}}
% \affiliation{Universidad de la Rep\'ublica, Facultad de Ciencias, Igu{\'a} 4225, Montevideo, Uruguay.}
% 
% \author{Cecilia Cabeza}
% \affiliation{Universidad de la Rep\'ublica, Facultad de Ciencias, Igu{\'a} 4225, Montevideo, Uruguay.}
%%%%%%%%%%%%%%%%%%%%%%%%%%%%%%%%%%%%%%%%%%%%%%%%%%%%%%%%%%%%%%%%%%%%%%%%%%%%%%%%%%%
\date{\today}

\abstract{The logistic map is a paradigmatic dynamical system originally conceived to model the discrete-time 
demographic growth of a population, which shockingly, shows that discrete chaos can emerge from trivial 
low-dimensional non-linear dynamics. In this work, we design and characterize a simple, low-cost, 
easy-to-handle, electronic implementation of the logistic map. In particular, our implementation allows 
for straightforward circuit-modifications to behave as different one-dimensional discrete-time systems. 
Also, we design a coupling block in order to address the behavior of two coupled maps, although, our 
design is unrestricted to the discrete-time system implementation and it can be generalized to handle 
coupling between many dynamical systems, as in a complex system. Our findings show that the isolated 
and coupled maps' behavior has a remarkable agreement between the experiments and the simulations, 
even when fine-tuning the parameters with a resolution of $\sim 10^{-3}$. We support these conclusions
by comparing the Lyapunov exponents, periodicity of the orbits, and phase portraits of the numerical 
and experimental data for a wide range of coupling strengths and map's parameters.
\PACS{05.45.-a 
         \and
      07.05.Fb
      \and 
      07.50.Ek
     } % end of PACS codes
}

\maketitle
% As a general rule, do not put math, special symbols or citations in the abstract or keywords.
%\begin{abstract}

\keywords{logistic map \and coupling \and periodicity}
%\end{abstract}
%\pacs{}
%\keywords{coupling}


%%%%%%%%%%%%%%%%%%%%%%%%%%%%%%%%%%%%%%%%%%%%%%%%%%%%%%%%%%%%%%%%%%%%%%%%%%%%%%%%%%%
\section{Introduction}
Nowadays, there is a growing scientific interest in explaining the collective behaviors that emerge in complex 
systems \cite{Barabasi_2002,Strogatz_2003,Barrat_2008}, namely, systems composed of many non-linearly 
interacting subsystems. However, the analysis of complex systems is usually restricted to \textit{toy-models} 
or numerical experiments \cite{lorenz1962,winfree_2001,kaneko1.4916925,Lloyd_1995,kendall1998spatial}, 
discarding the intractable parameter heterogeneities and random fluctuations (caused by intrinsic noise sources) 
found in real-world complex systems. Hence, the analysis of a synthetic complex system from a versatile experiment 
is always well-posed.

A well-known paradigmatic model that is used to understand chaotic behavior emerging from a trivial non-linear evolution
is the logistic map \cite{may1976simple,feigenbaum_1978,Yorke1996,collet2009iterated}. It constitutes a tractable mathematical 
benchmark to characterize chaotic behavior (and other emerging phenomena) with a vast range of applications. For example, 
it has been used as a noise generator \cite{mcgonigal1987generation,phatak_1995}, a simple ecological model \cite{stone1993period}, 
an encryption machine for secure communications \cite{pareek2006image,singh2010chaos,borujeni2015modified}, or even introducing 
extensions to include more degrees of freedom \cite{baptista_1996,baptista_1997,campos2011family,radwan2013some}. Similarly, 
the inclusion of coupling between maps \cite{kaneko_1983,kaneko1990clustering,maistrenko_1998}, for example, as in ecological
models that address the effects of diversity or spatial heterogeneity in competing populations \cite{Lloyd_1995,kendall1998spatial},
shows promising results and increases the degrees of freedom. Analogously, a recent experimental study of coupled oscillators 
\cite{PhysRevE.92.052912} shows the increase in complexity due to the coupled dynamics. Nevertheless, the interacting model 
of logistic maps still provides a solid benchmark to address emerging behaviors in other complex systems 
\cite{collet2009iterated,kaneko1.4916925}. 

In this work, we design a versatile experimental implementation of a complex system composed of interacting logistic maps.
Former attempts to electronically implement a logistic map are scarce and lack in simplicity 
\cite{horowitz1989art,suneel2006electronic,garcia2013difference}, although manage to maintain control over all parameter range. 
Our logistic map design is simple and in a block form, has low-cost electronic components, it is easy-to-handle 
(allowing to maintain control over parameters), and also shows low power-consumption. Moreover, it allows to modify the map's 
block to include different behaviors, specifically, to become a different one-dimensional discrete-time system or even a 
continuous-time version of the system. The interaction between maps is designed as a coupling block that addresses the behavior 
of two coupled maps in the Kaneko style \cite{kaneko1990clustering}, however, this block is unrestricted to our case-study. 
In general, our coupling block can handle couplings between continuous-time dynamical systems and it also allows for 
straightforward extensions to many interacting dynamical systems. Such extensions make our model a versatile option to 
experimentally study different complex systems' behavior.

Our experimental findings show remarkable agreement with all numerical experiments, which we corroborate by exploring a 
wide range of parameters using high resolution ($\sim 10^{-3}$). Specifically, we explore variations in the logistic map's
parameter and coupling strength for the uncoupled and coupled situations, respectively. Hence, we characterize the bifurcation
cascades that the uncoupled and coupled systems exhibit \cite{feigenbaum_1978} in both, numerical and experimental data, with 
high accuracy and signal-to-noise ratio. The comparative analysis between experiments and simulations shows that the level of 
performance and agreement is excellent. Moreover, we support this by analyzing the numerical and experimental data using Lyapunov 
exponents \cite{Yorke1996}, orbit's periodicity \cite{cromer_1989}, and phase portraits for the uncoupled and coupled cases.

The present work is organized as follows. We commence in Sec.~\ref{sec:model} by introducing the model of coupled logistic
maps and the methods used in the subsequent sections for the analysis of the experimental and numerical data. The experimental 
setup is presented in Sec.~\ref{sec:elec}, starting with the electronic implementation of a single logistic map and the 
description of the coupling block and ending with the explanation on how to obtain an experimental discrete-time coupled system. 
The characterization of the dynamics of a single and a pair of maps is analyzed in Sec.~\ref{sec:res} by means of bifurcations 
diagrams and stability charts displaying the periodicity (or chaoticity) as a function of the parameters. We end in Sec.~\ref{sec:con} by summarizing our main conclusions.

%%%%%%%%%%%%%%%%%%%%%%%%%%%%%%%%%%%%%%%%%%%%%%%%%%%%%%%%%%%%%%%%%%%%%%%%%%%%%%%%%%%
 \section{Model and Methods} \label{sec:model}

%%%%%%%%%%%%%%%%%%%%%%%%%%%%%%%%%%%%%%%%%%%%%%%%%%%%%%%%%%%%%%%%%%%%%%%%%%%%%%%%%%%
 \subsection{Coupled logistic maps}
The celebrated logistic map describes the discrete-time evolution
of a closed population \cite{may1976simple,feigenbaum_1978,Yorke1996,collet2009iterated},
\begin{equation}
 x_{n+1}=  f\left(r,x_{n}\right) = r\,x_{n}\left(1-x_{n}\right),
 \label{eq:eqLogistico}
\end{equation}
where $x_n$ represents the ratio of the population to a maximum value ($x_n \in [0,1]$) at time $n$ (discrete) 
and $r$ is the control parameter, which is restricted to the interval $(0,4]$ in order to keep the normalized population in 
the interval $[0,1]$. The nonlinear term in the right hand side of Eq.~(\ref{eq:eqLogistico}) accounts for the population growth 
by reproduction (e.g., when the population size is small, $x_{n+1} \sim r\,x_n$) and for the starvation due to the limit imposed 
by the carrying capacity of the environment ($x_{n+1} \sim r[1-\,x_n]$).

In this work, aside from the experimental implementation and characterization of the isolated logistic map 
(Eq.~\ref{eq:eqLogistico}), we analyze two coupled maps. Hence, for this case the state variables are
$x_{n}^{(1)}$ and $x_{n}^{(2)}$, with control parameters $r_1$ and $r_2$, respectively. The evolution of 
these coupled maps is determined from
\begin{equation}
\left\{ \begin{array}{c}
x_{n+1}^{(1)}=\left(1-\epsilon\right)f\left(r_{1},x_{n}^{(1)}\right)+\epsilon f\left(r_{2},x_{n}^{(2)}\right),\\
x_{n+1}^{(2)}=\left(1-\epsilon\right)f\left(r_{2},x_{n}^{(2)}\right)+\epsilon f\left(r_{1},x_{n}^{(1)}\right),
\end{array}\right.\label{eq:SystemaEcs}
\end{equation}
where $\epsilon$ represents the coupling strength. This approach can be generalized to a network of $N$ maps 
\cite{kaneko1990clustering} by
\begin{equation}
x_{n+1}^{(i)}=\left(1-\epsilon \right)f\left(r_{i},x_{n}^{(i)}\right)+\epsilon\sum_{j=1}^{N}\frac{A_{ij}}{d_{i}}f\left(r_{j},x_{n}^{(j)}\right),\label{eq:acople Kaneko}
\end{equation}
where $A_{ij}$ is the $ij$ entry of the adjacency matrix of the network and $d_i = \sum_j A_{ij}$ is the $i$-th node degree.

%%%%%%%%%%%%%%%%%%%%%%%%%%%%%%%%%%%%%%%%%%%%%%%%%%%%%%%%%%%%%%%%%%%%%%%%%%%%%%%%%%%
 \subsection{Dynamical behavior characterization}
Lyapunov exponents provide a useful characterization of a dynamical system in terms of how sensible the system is to 
small changes in its initial conditions \cite{Yorke1996}. Specifically, they quantify the average divergence of an 
infinitesimal displacement from an unperturbed reference orbit and are related to the factor by which the infinitesimal 
displacement grows or shrinks. For a generic dynamical system, the number of Lyapunov exponents is equal to the dimension 
of the system, i.e., the number of independent perturbations. In many applications it is sufficient to calculate only the 
largest Lyapunov exponent, $\lambda_{max}$, since in general, $\lambda_{max} > 0$ ($\lambda_{max} < 0$) implies the presence 
of chaotic (periodic) behavior.

In the case of an isolated logistic map, namely, a one-dimensional discrete-time dynamical system, there is a single Lyapunov
exponent, $\lambda$, to determine. This exponent is directly obtained from \cite{Yorke1996}
\begin{equation}
\lambda\left(r\right) = \lim_{N\to\infty}\frac{1}{N}\sum_{n=1}^{N-1}\ln\left[r\,\left( 1 - 2\,x_{n}\right)\right],
 \label{eq:Lyap_exp}
\end{equation}
where $r$ is the map's parameter (Eq.~\ref{eq:eqLogistico}) and $x_n$ for $n = 1$ corresponds to the initial condition of 
the map. For finite-size orbits, namely, when the limit is absent, $\lambda$ is a finite-time Lyapunov exponent (FLE), which 
we name in what follows as Lyapunov exponent. However, for sufficiently long time-series ($N\gg1$), the FLE converges 
asymptotically to the value of (Eq.~\ref{eq:Lyap_exp}).

Another useful characterization for the coupled maps behavior is done by quantifying the periodicity of their orbits 
\cite{cromer_1989}. In other words, looking at the periodic properties of the resultant orbits for every dynamical regime.
Specifically, we measure the periodicity of an orbit as a function of the parameters for the coupled system. In order to 
measure this periodicity, we define the period of the coupled system by building a data sequence that is defined by 
concatenating the state variables at consecutive times. For example, in our case-study, the coupled maps orbit results in 
a concatenated time-series as
\begin{equation}
 \ldots x^{(1)}_n,x^{(2)}_n,x^{(1)}_{n+1},x^{(2)}_{n+1}, \ldots 
 \label{eq:periodicity}
\end{equation}
Consequently, the period of the system is identified by the periodicity of the compound state sequence divided by the 
number of units, which in this work is $2$.

%%%%%%%%%%%%%%%%%%%%%%%%%%%%%%%%%%%%%%%%%%%%%%%%%%%%%%%%%%%%%%%%%%%%%%%%%%%%%%%%%%%
\section{Electronic Implementation}
\label{sec:elec}
Our logistic map design is divided into two main parts. A logistic-map block (LMB), which makes an analog logistic 
function [as in the right-hand side of Eq.~\ref{eq:eqLogistico}], and a sample-and-hold block (SHB), which samples 
the voltage of the continuously varying analog-signal of the LMB and holds its value at a constant level for a specified 
period of time. Hence, the output is an analog time-series that varies its values step-wise, modeling the discrete evolution 
of a map. This electronic implementation provides great flexibility since it allows to design other maps by modifying the LMB 
or to implement time-delayed models (as in Refs.~\cite{amil2015electronic,amil2015organization}) by modifying the SHB. Furthermore,
this design is chosen for scalability, meaning that its implementation allows for the direct introduction of coupling between 
several individual units and with arbitrary connections among them. In other words, our implementation in blocks allows to set 
different networks of coupled maps without the commonly found limitation of an increasing complexity in the electronic setup. 
In particular, the coupling between two logistic maps as in Eq.~\ref{eq:SystemaEcs} is designed in a similar block form, which 
we name the coupling block (CB). Hence, we retain the scalability of the system allowing for a possible extension of the design 
to contemplate a coupled $N$-maps dynamic [Eq.~\ref{eq:acople Kaneko}].

%%%%%%%%%%%%%%%%%%%%%%%%%%%%%%%%%%%%%%%%%%%%%%%%%%%%%%%%%%%%%%%%%%%%%%%%%%%%%%%%%%%
\subsection{The Logistic-Map Block}

\begin{figure}
\includegraphics[width=0.99\columnwidth]{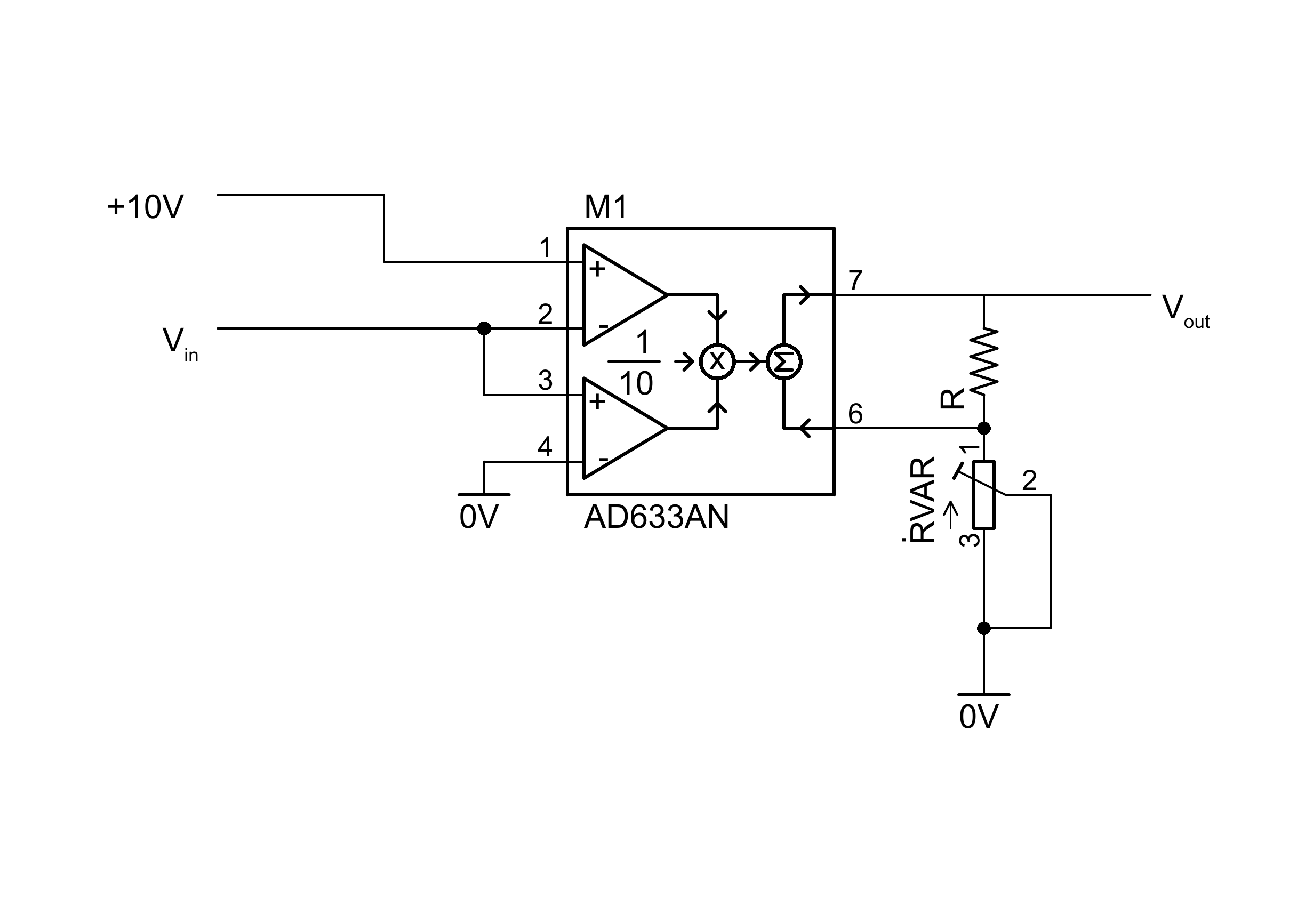}
\caption{Diagram of the electronic circuit that reproduces the logistic function. Namely, the output voltage is a quadratic 
function of the input voltage.
\label{fig:Logistic-block-diagram}}
\end{figure}

The LMB is designed to reproduce the logistic function of the right-hand side of Eq.~\ref{eq:eqLogistico} and is shown
in Fig.~\ref{fig:Logistic-block-diagram}. The present implementation uses an analog multiplier AD633, whose input range is 
$\pm10V$ and output range $\pm11V$. The output voltage, $V_{out}$, is obtained using the information provided by the manufacturer,
namely,
\begin{equation}
V_{out}(t) = \frac{ [ V_{1}(t) - V_{2}(t) ]\,[ V_{3}(t) - V_{4}(t) ] }{ V_{s} } + V_{6}(t),
 \label{eq_first_LMB}
\end{equation}
$V_i(t)$ ($i = 1,\ldots,6$) being the voltage at the terminals indicated in Fig.~\ref{fig:Logistic-block-diagram} and 
$V_{s}=10V$ being the saturation voltage of the AD633. For this circuit, assuming ideal behavior of the components and 
applying Kirchhoff's laws \cite{horowitz1989art}, we find from Eq.~\ref{eq_first_LMB} that
\begin{equation}
 V_{out}(t) = \left( 1 + \frac{R_{var}}{R} \right) \frac{V_{in}(t) \,( V_{s} - V_{in}(t) )}{ V_{s} },
 \label{eq:function}
\end{equation}
where $R = 1k\Omega$ and $R_{var}$ can be set between $0 \Omega$ and $3k\Omega$. 
The different values of $R_{var}$  are obtained using a step-by-step motor attached to a multi-turn potentiometer and 
controlled by a National Instrument Data Acquisition (NIDAQ). 

The voltages of the electronic circuitry are identified with the state-variables of Eq.~\ref{eq:eqLogistico} by 
$x_{n} \mapsto V_{in}/V_{s}$ and $x_{n+1} \mapsto V_{out}/V_{s}$, and the control parameter with
\begin{equation}
r \equiv \left(1+\frac{R_{var}}{R}\right).
\label{eq:r_exp}
\end{equation}

Our analysis of the LMB takes into account solely the variation of $r$ in the interval $(1,4)$. The reason is that, 
since $R = 1k\Omega$ under our implementation, the interval $r\in(0,1)$ where the dynamics of the map corresponds to a 
stable fixed-point is unreachable. Also note that, when $r \simeq 4$, the electronic noise can take the voltage to values 
higher than $V_{s}$, hence, saturating the analog multiplier. Consequently, in the analysis to characterize the system we 
set $r < 4$ such that we always avoid this problem.

%%%%%%%%%%%%%%%%%%%%%%%%%%%%%%%%%%%%%%%%%%%%%%%%%%%%%%%%%%%%%%%%%%%%%%%%%%%%%%%%%%%
\subsection{The Sample-and-Hold Block}

\begin{figure}[htbp]
\includegraphics[width=0.99\columnwidth]{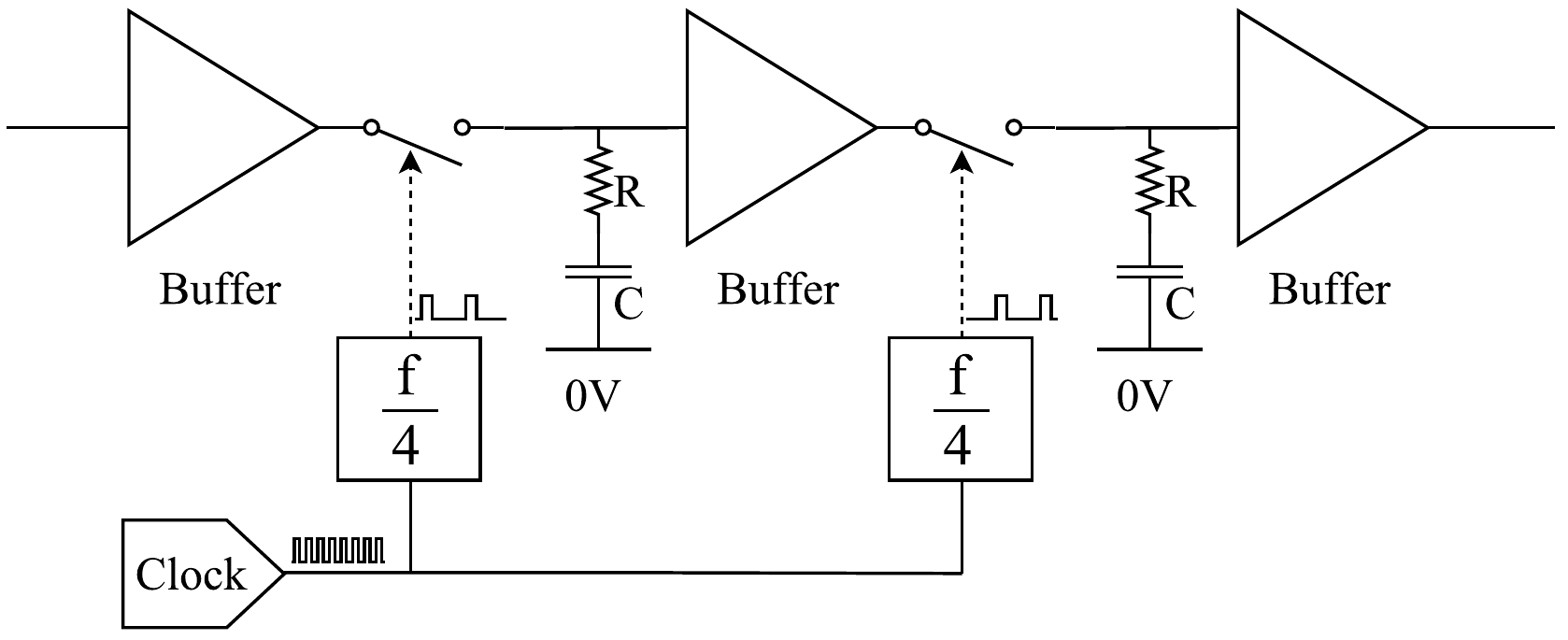}
\caption{
\label{fig:Sample-and-Hold}
Schematic diagram of the sample-and-hold block
. This block is implemented to produce a step-wise evolution of the 
electronic circuit, hence, a discrete-time evolution is achieved.}
\end{figure}

The SHB is based on two LF398 circuits schematically shown in Fig.~\ref{fig:Sample-and-Hold} \cite{suneel2006electronic}. 
It samples the voltage from the input terminal at an instant of time, keeps its value in the hold capacitor, and then releases
its value from the output terminal one clock-period later. Every two clock periods, the roles of both LF398 are interchanged. 
This switching results in a discontinuous evolution of the whole circuit (LMB plus SHB), where at each instant of time a value 
of $x_{n}$ is obtained.

The optimal clock's frequency, which sets the time lapse between consecutive values of the output voltage, must be chosen taking 
into account several experimental constrains. On the one hand, there is a limit given by the time it takes for the SHB to charge 
the capacitors. On the other hand, the existence of parasitic capacitance, bias currents in the operational amplifiers or other 
components, and leakage currents in the capacitors, also sets limits to the clock's frequency. Moreover, the response time for 
the rest of the circuit to stabilize after any change, i.e., the time needed by the LMB and the coupling to stabilize the output, 
constitutes an upper bound for the clock's frequency. However, from a practical point of view, the clock's frequency should be as 
high as possible to reduce the time necessary to perform the experiments and obtain long time-series. Consequently, we have chosen 
the clock's frequency to be in the range between $10kHz$ to $20kHz$.

%%%%%%%%%%%%%%%%%%%%%%%%%%%%%%%%%%%%%%%%%%%%%%%%%%%%%%%%%%%%%%%%%%%%%%%%%%%%%%%%%%%
\subsection{The coupling block}

\begin{figure}[htbp]
\includegraphics[width=0.9\columnwidth]{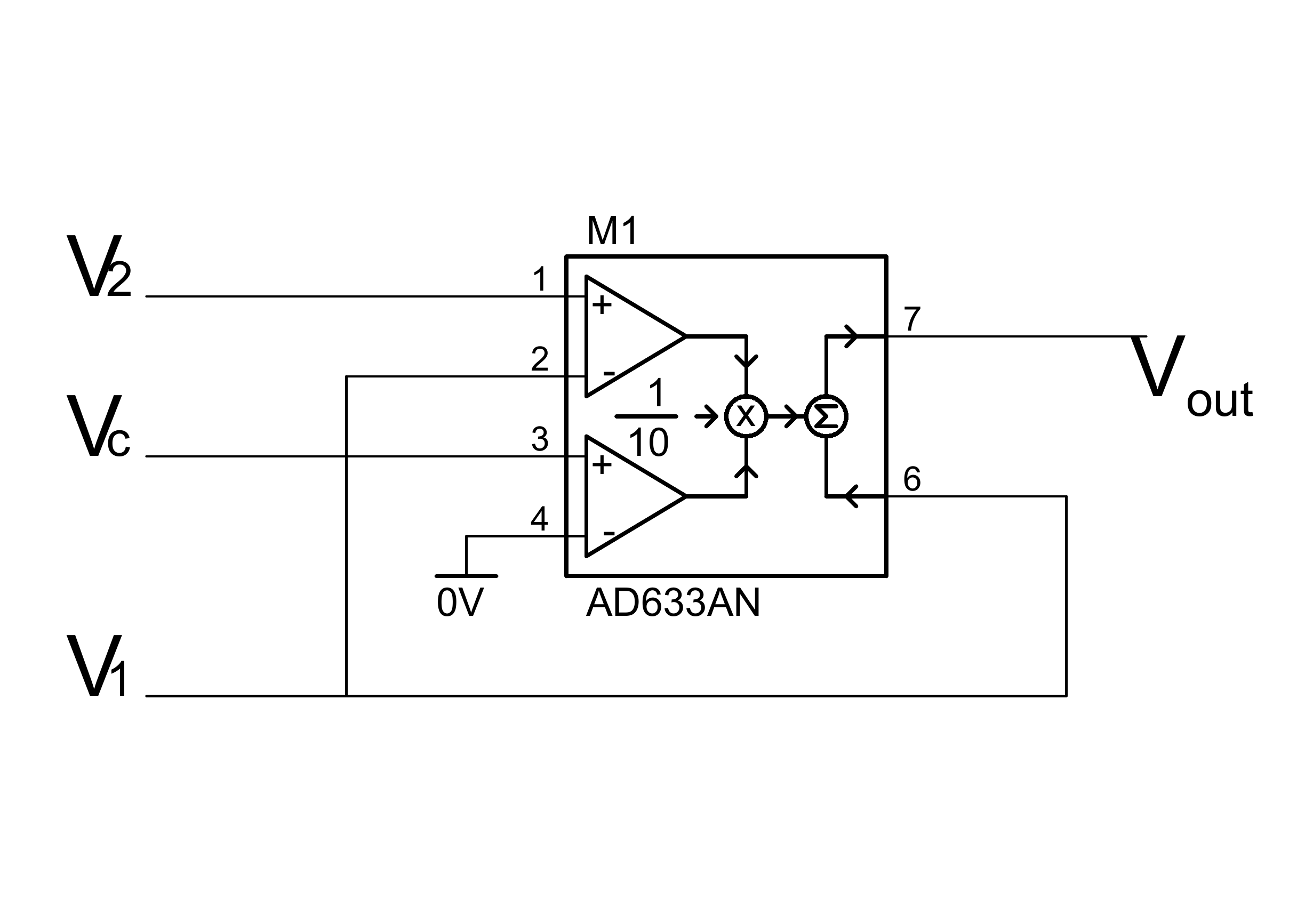}
\protect\caption{\label{fig:Coupling-Block}
Diagram of the coupling circuit. The output voltage from this circuit linearly relates both input voltages, 
$V_1$ and $V_2$, with a coupling intensity given by $V_c$.}
\end{figure}

\begin{figure}[htbp]
\includegraphics[width=0.95\columnwidth]{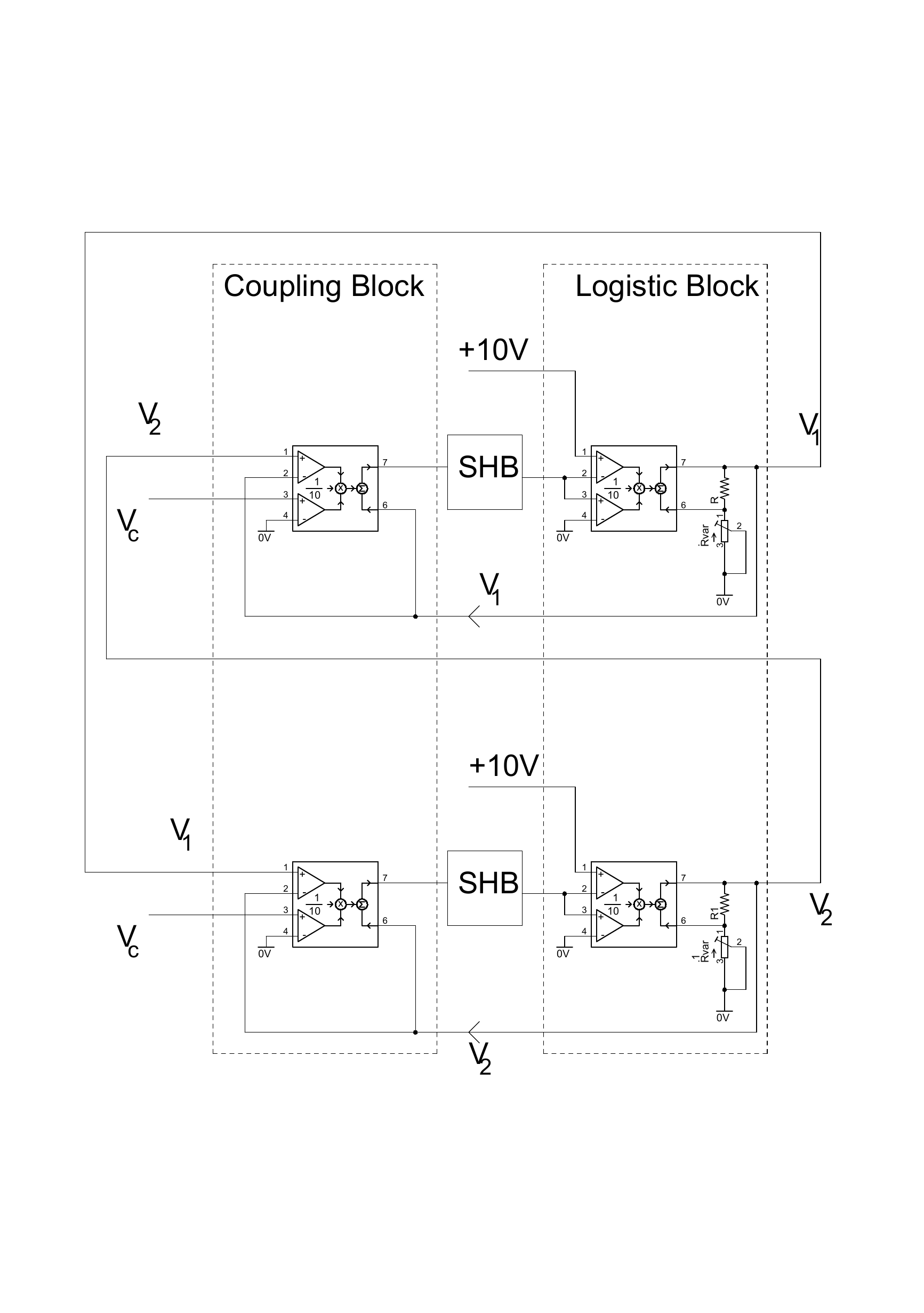}
\protect\caption{\label{fig:Block-diagram-coupled-logistic}
Block diagram of two coupled logistic maps. SHB stands for the sample-and-hold block of Fig.~\ref{fig:Sample-and-Hold}.}
\end{figure}

The experimental setup for the coupling circuit is depicted in Fig.~\ref{fig:Coupling-Block}. In order to couple two 
logistic maps, we require a coupling circuit for each map. Moreover, to retain the possibility of extending our design to 
implement networks of maps, we design a coupling block (CB) as in Fig.~\ref{fig:Block-diagram-coupled-logistic}. Hence, the
LMB is connected to the SHB to define the discrete-time evolution of the system and it is also connected to the CB to implement 
the coupled evolution according to Eq.~\ref{eq:SystemaEcs}. Specifically, after taking into account the AD633 in 
Fig.~\ref{fig:Coupling-Block}, we obtain $V_{out}$ as
\begin{equation}
 V_{out}(t) =  \left(1 - \frac{ V_{c} }{ V_{s} }\right) V_{1}(t) + \frac{ V_{c} }{ V_{s} } V_{2}(t),
\end{equation}
where $V_{c}$ is the control voltage for our CB, namely, the coupling strength between the maps; $\epsilon \equiv V_c/V_s$. 
In particular, when $V_{c}$ changes between $0$ and $V_{s}$ (the saturation voltage), $\epsilon$ changes between $0$ and $1$. 
Thus,
\begin{equation}
 x_{out}(t) = \frac{V_{out}(t)}{V_s} = \left(1 - \epsilon \right)\,x_{1}(t) + \epsilon\,x_{2}(t),
\end{equation}
where we have identified the voltages $V_1(t)$ and $V_2(t)$ with the corresponding logistic states 
$x^{(1)}(t) = V_1(t)/V_s$ and $x^{(2)}(t) = V_2(t)/V_s$, respectively.

In order to have high accuracy and control over the changes in $V_c$ we use the analog output of the NIDAQ, 
which allows to set $V_{c}$ with a precision of $\pm20mV$ (according to the data-sheet). This precision is even lower
than the $30mV$ noise level we observe experimentally in our time-series data. Moreover, we tested the excellent 
performance of the CB by critically comparing the experimental time-series with numerically generated time-series of 
Eq.~\ref{eq:SystemaEcs}.

%%%%%%%%%%%%%%%%%%%%%%%%%%%%%%%%%%%%%%%%%%%%%%%%%%%%%%%%%%%%%%%%%%%%%%%%%%%%%%%%%%%
\subsection{Electronic map}
\begin{figure}[htbp]
 \includegraphics[width=0.95\columnwidth]{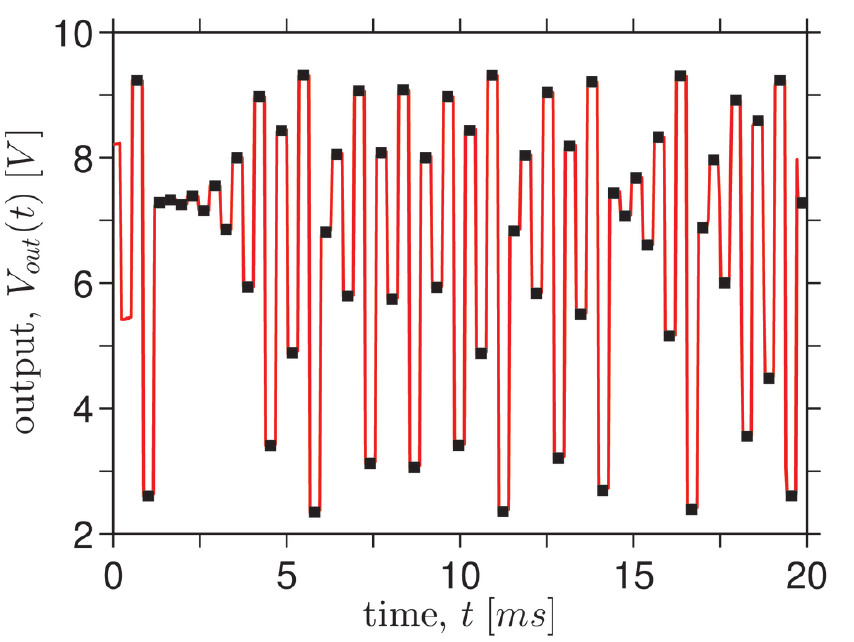}
\protect\caption{\label{fig:Logistic-map-time-peaks}
Experimental output voltage obtained from our logistic-map block (LMB) implementation with its sample-and-hold block (SHB). 
The analog output signal is represented by the continuous line and the discrete map evolution that is obtained from this output is
represented by the filled squares. The data corresponds to a single LMB plus SHB for the case where there is coupling between two 
of these circuits with a strength $\epsilon=0.5$ and the circuit's parameters are in the chaotic regime ($r=3.8$).}
\end{figure}

The discrete time-evolution of the experimental system is obtained after processing the continuous step-wise evolution 
of the LMB plus SHB output. The conversion from the combined circuit's output signal to the discrete-time state variable of 
the logistic map, namely, $x_n$, consists in taking the mean value for each plateau. Specifically, $x_n$ is found by the hold 
value of the sampling of the LMB voltage every four clock-periods. A working example of this process is shown in
Fig.~\ref{fig:Logistic-map-time-peaks}, where the continuous signal registered from the LMB plus SHB is represented by 
the continuous line and the corresponding discrete evolution is represented by the signaled square points. As a result, 
the storage space that is needed to save the output signal is reduced and we obtain a discrete-time evolution of the system, i.e., 
the logistic map's evolution.

In particular, our findings show that there is a remarkable agreement between the experimental values of $r$, which are found 
from Eq.~\ref{eq:r_exp}, and the $r$ we found from fitting this time-series output to a quadratic function. For example, 
the output of the LMB using an experimental control parameter of $r = 3.5\pm 0.1$ [Eq.~\ref{eq:r_exp}], results in a time-
series with a fitted value of $r = 3.5005$ and regression coefficient of $0.9999$.

%%%%%%%%%%%%%%%%%%%%%%%%%%%%%%%%%%%%%%%%%%%%%%%%%%%%%%%%%%%%%%%%%%%%%%%%%%%%%%%%%%%
\section{Results and Analysis}
\label{sec:res}
%%%%%%%%%%%%%%%%%%%%%%%%%%%%%%%%%%%%%%%%%%%%%%%%%%%%%%%%%%%%%%%%%%%%%%%%%%%%%%%%%%%
\subsection{Isolated logistic map}
To corroborate that our electronic model reproduces correctly the logistic map's behavior, the experimental results are compared
with the numerical simulations of Eq.~\ref{eq:eqLogistico}. All comparisons are performed by neglecting a transient of $10^{4}$ 
iterations, which eliminates the orbit's dependency on the initial condition.

\begin{figure}[htbp]
\includegraphics[width=0.95\columnwidth]{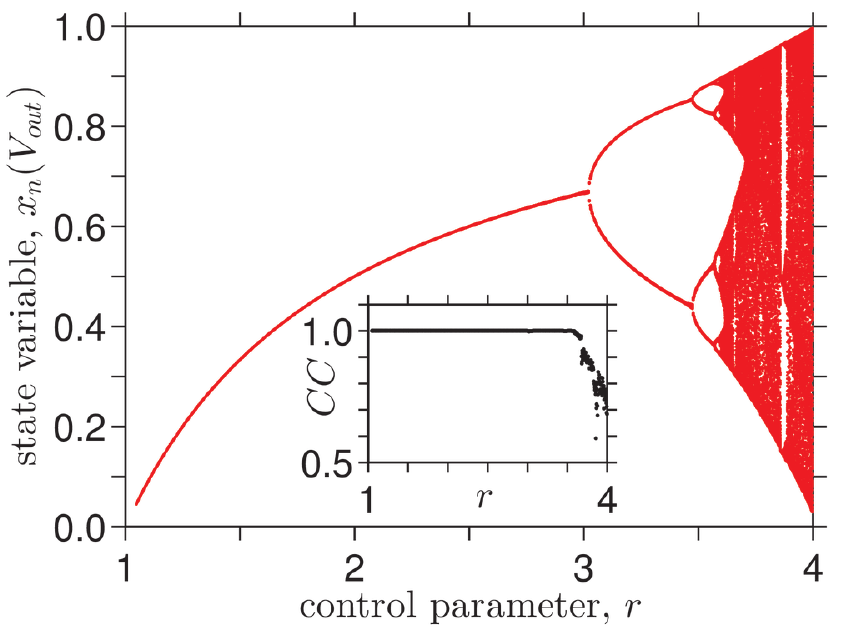}
 \protect\caption{\label{fig:Bifurcation-diagram-:}
Bifurcation diagram for the output voltages of our logistic map's electronic implementation. The diagram is
constructed discarding $10^4$ initial time-iterations and taking the next $256$ values of the discrete-time voltages 
($x_n$) as a function of $1024$ different control parameter ($r$) values. It shows the distinctive traces of the 
logistic map's route to chaos. The correlation coefficients ($CC$) between the experimental data and the numerical 
simulation for each $r$ are shown in the inset, where small ($<30\%$) deviations are found for parameter values close 
to the chaotic region ($r > 3.5$).}
\end{figure}

The experimental bifurcation diagram as a function of the control parameter is shown in Fig.~\ref{fig:Bifurcation-diagram-:}. 
As it is seen, it reproduces the essential characteristics of the logistic map's Feigenbaum diagram \cite{feigenbaum_1978,Yorke1996}.
The agreement is quantified by the high value that the correlation coefficient, $CC$, between the experimental 
and the simulated signal have, which is shown in the inset of Fig.~\ref{fig:Bifurcation-diagram-:}, specially for
the periodic regions. In particular, the slight departures of $CC$ from unity for $r > 3.5$ are a consequence of 
chaos and small shifts in the experimental control parameter value. However, as it is seen from the figure, these 
shifts leave the diagram virtually unaffected with respect to the Feigenbaum diagram of the logistic map 
\cite{feigenbaum_1978,Yorke1996}.
 
\begin{figure}[htbp]
 \includegraphics[width=0.95\columnwidth]{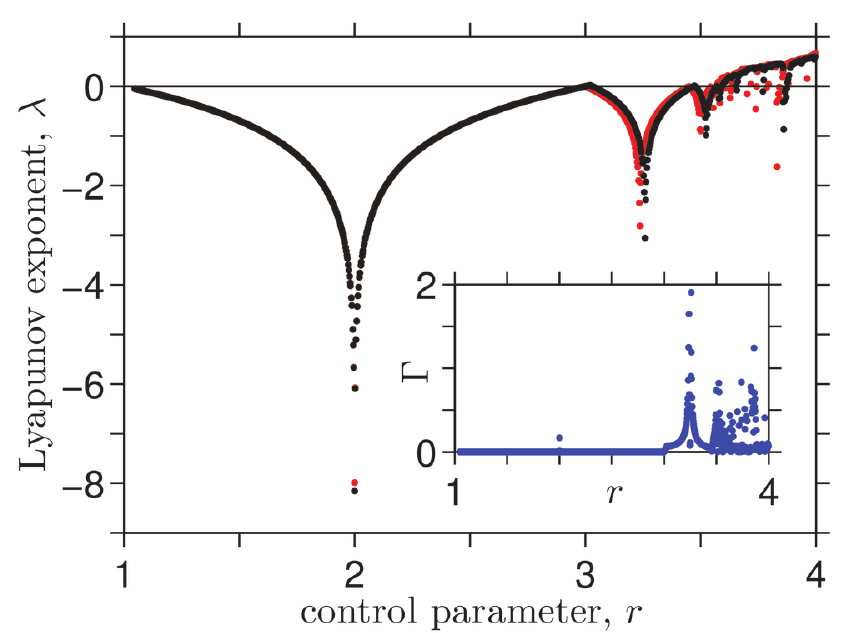}
 \protect\caption{\label{fig:Lyapunov-exponents-:}
(Color online) Experimentally (black online) and numerically (red online) obtained Lyapunov exponents ($\lambda$) for a 
logistic map. Both exponents are found from $10^4$ iterations for $1024$ different control parameter ($r$) values. The differences 
between these exponents for each value of $r$ are quantified in the inset by the error function, 
$\Gamma \equiv |\lambda_{sim} - \lambda_{exp}|$, where $\lambda_{exp}$($\lambda_{sim}$) is the experimental (numerical) 
Lyapunov exponent for the particular time-series.}
\end{figure}

In order to further analyze the dynamical behavior of our electronically implemented map, we compare the Lyapunov exponents
of the experimentally and numerically obtained time-series. In particular, the experimental Lyapunov exponents are calculated
from $N = 10^4$ data points, instead of the $N = 256$ data points that are used to construct the experimental diagram of 
Fig.~\ref{fig:Bifurcation-diagram-:}. Similarly, the numerical simulations are iterated $N = 10^4$ times. The resultant 
Lyapunov exponents for both cases are shown in Fig.~\ref{fig:Lyapunov-exponents-:}. The Pearson's correlation coefficient 
between the experimental and the simulated Lyapunov exponents we find is $CC = 0.9816$. Moreover, we calculate the error 
between the exponents, $\Gamma \equiv |\lambda_{exp}-\lambda_{sim}|$, for each parameter $r$ value, which is shown in the 
inset of Fig.~\ref{fig:Lyapunov-exponents-:}. $\Gamma$ shows, as the bifurcation diagram does, that there is a remarkable 
agreement between the experimental and the numerical data in the regions displaying periodic behavior. The discrepancies 
that are found in the chaotic region correspond to a small shift in the experimental control parameter, $r$, which looks as a 
rigid body translation of the experimental system with respect to the simulation. Nevertheless, these analysis reveals that 
our electronic implementation reproduces the behavior of the logistic map with high accuracy, specially for $r < 3$.

%%%%%%%%%%%%%%%%%%%%%%%%%%%%%%%%%%%%%%%%%%%%%%%%%%%%%%%%%%%%%%%%%%%%%%%%%%%%%%%%%%%
\subsection{Two coupled maps}

Here we present the experimental and numerical results for the coupled dynamics of two logistic-maps.

\begin{figure}[htbp]
 \begin{center}
%  \text{\bf (a)}\\
\includegraphics[width=0.95\columnwidth]{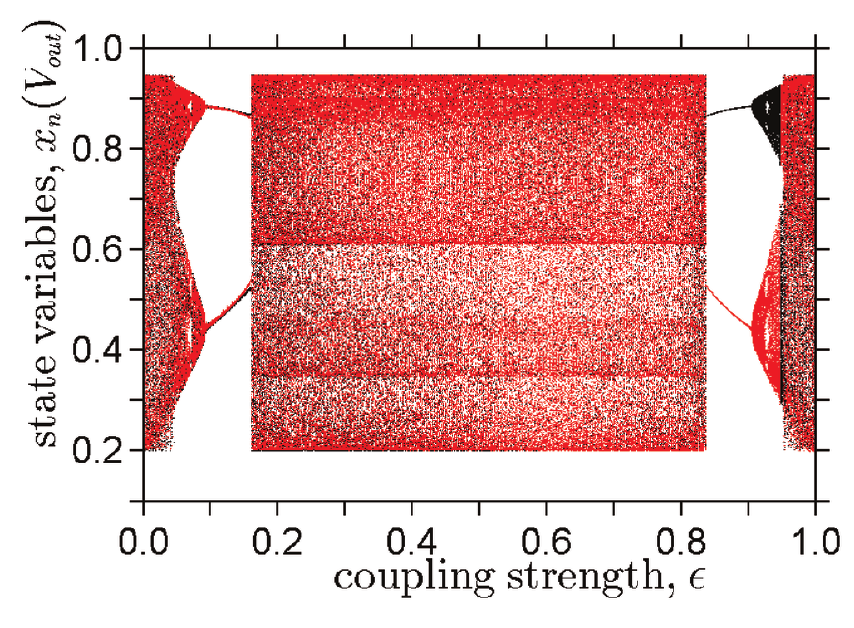} \\
 % \text{\bf (b)}\\
 \includegraphics[width=0.95\columnwidth]{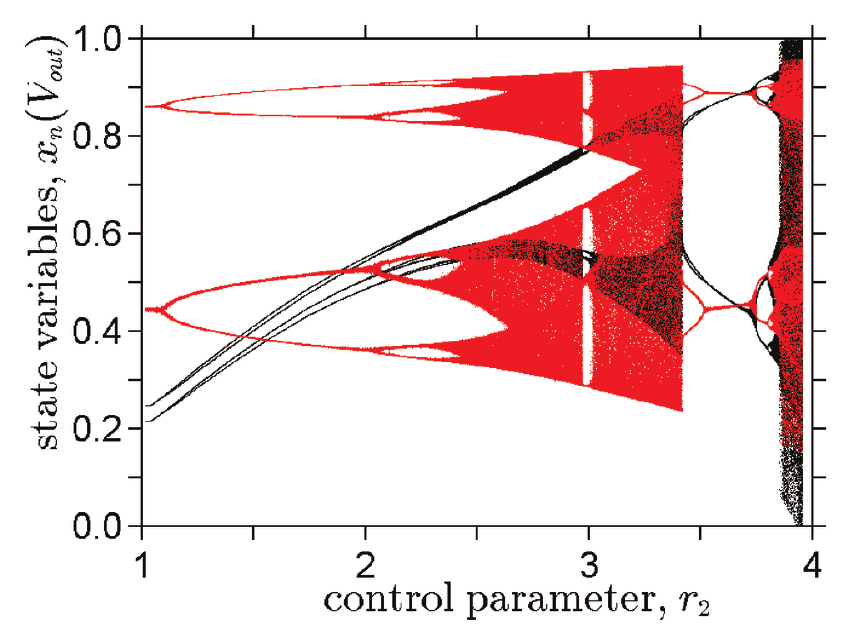}
 \end{center}\vspace{-1pc}
 \protect\caption{\label{fig:Coupled-maps-bif}
(Color online) Experimental bifurcation diagrams. Panel {\bf (a)} is obtained from increasing the coupling strength 
between two identical maps, where both logistic maps are in the chaotic region ($r=3.8$) and the increments in $\epsilon$ 
are set to $0.01$. Panel {\bf (b)} is obtained by decreasing the heterogeneity between the maps, namely, 
$r_1 = 3.8$, $\epsilon = 0.1$, and we increase $r_2$. The light (red online) and dark (black online) dots on both panels 
correspond to the time-series from the maps $x_n^{(1)}$ and $x_n^{(2)}$, respectively.}
\end{figure}

The bifurcation diagrams for identical maps as a function of the coupling strength, $\epsilon$, are displayed in 
Fig.~\ref{fig:Coupled-maps-bif}{\bf (a)}, where the map's parameters are set equally to $r_1 = r_2 = r = 3.8$. Due to the 
presence of coupling between the maps, namely, $\epsilon > 0$, the resultant dynamics shows both chaotic (regions where points
spread vertically in Fig.~\ref{fig:Coupled-maps-bif}{\bf (a)}) and periodic (thin lines of points in
Fig.~\ref{fig:Coupled-maps-bif}{\bf (a)} that stay narrow over a range of $\epsilon$ values) behavior. These behavior contrasts 
the exclusively chaotic behavior that the maps show for this parameter value ($r = 3.8$) in the absence of coupling, as it is
seen on the left of Fig.~\ref{fig:Coupled-maps-bif}{\bf (a)} for $\epsilon = 0$. Two videos for the bifurcation development 
are presented on the Supplementary Material (SM) corresponding to the weak (Fig.~\ref{fig:Coupled-maps-bif}{\bf (a)} left) and
strong (Fig.~\ref{fig:Coupled-maps-bif}{\bf (a)} right) coupling scenarios. These videos show the phase space portraits as 
coupling $\epsilon$ is increased from $0$ to $0.12$ ($0.82$ to $1$), hence, changing the chaotic (periodic) behavior and shifting
it to a periodic (chaotic) state. Also, the videos show how quasi-periodic behavior emerges before periodic or chaotic behavior does.
%[Pablo]: Cuasi-periódicos

A non-symmetrical situation is depicted in the bifurcation diagram of Fig.~\ref{fig:Coupled-maps-bif}{\bf (b)}.
%In this bifurcation diagram, the transformation $\epsilon \mapsto 1  - \epsilon$ and $x_{1,2} \mapsto x_{2,1}$ is violated since
We are fixing one map's parameter ($r_1 = 3.8$) and the coupling strength ($\epsilon = 0.1$) but we are changing the other
map's parameter ($r_2$). Hence, this diagram is the quantification of how the decrease in heterogeneity between the maps 
(namely, the increase of $r_2$ from $1.0$ to $3.8$) causes bifurcations to emerge 
\cite{kaneko_1983,kaneko1990clustering,maistrenko_1998}. Hence, we shift from the periodic region 
(left side of Fig.~\ref{fig:Coupled-maps-bif}{\bf (b)}) to the chaotic region (right side of Fig.~\ref{fig:Coupled-maps-bif}{\bf (b)}). However, although $r_2 = 1$ in the periodic region, corresponding to a fixed-point state, due to the weak coupling between the maps ($\epsilon = 0.1$) it shows a period-$2$ behavior. This is also supported by a phase space portrait video we are presenting in the SM.

%\begin{figure}[htbp]
% \includegraphics[width=0.95\columnwidth]{fig_9}
% \protect\caption{\label{fig:Coupled-maps-bif2}
%(Color online) {\bf Experimental bifurcation diagram of two coupled maps as heterogeneity is decreased}. The control parameter of one logistic map is fixed to $r_1 = 3.8$ and the other logistic map, $r_2$, is increased from $1$ to $4$ (horizontal axis), decreasing the map's differences. The coupling strength between the maps is set to $\epsilon = 0.1$ and the colors follow the same pattern as in Fig.~\ref{fig:Coupled-maps-bif1}.}
%\end{figure}

\begin{figure}[htbp]
 \begin{center}
% \text{\bf (a)}\\
  \includegraphics[width=0.95\columnwidth]{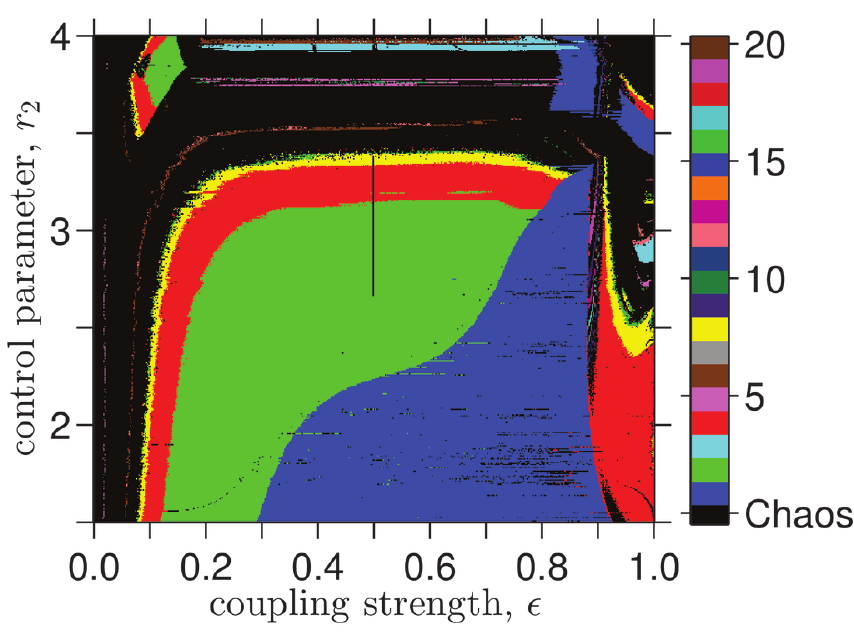}
 %\text{\bf (b)}\\
   \includegraphics[width=0.95\columnwidth]{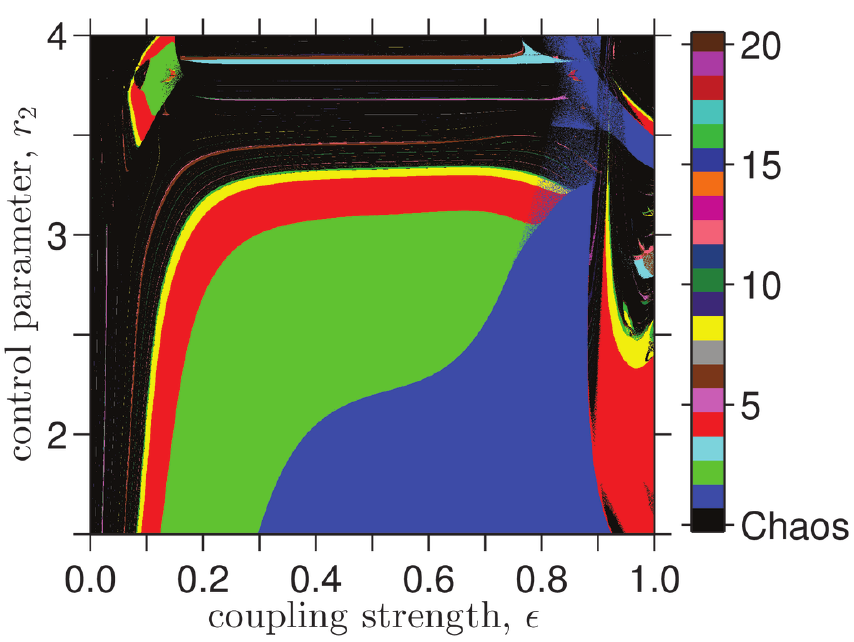}
 \end{center} \vspace{-1pc}
 \protect\caption{\label{fig:Coupled-maps-per}
 Orbit's periodicity for the parameter space of two coupled logistic-maps. The top (bottom) panel shows the experimental
(numerical) periodicity of 
the orbits in color scale. The parameter space is constructed by fixing $r_1 = 3.8$ and changing, $r_2$, and the coupling strength, 
$\epsilon$. Experimental (numerical) resolution: 512$\times$512  (2048$\times$2048) parameter points. }
\end{figure}

In order to continue the quantitative comparison between the experimentally implemented maps and the 
numerical simulations, we show in Fig.~\ref{fig:Coupled-maps-per} the periods that the orbits have as a 
function of the coupling strength ($\epsilon$, horizontal axes) and the map's control parameter ($r_2$, 
vertical axes), namely, the map's heterogeneity. Specifically, we fix $r_1 = 3.8$ and vary $\epsilon$ and $r_2$. 
These results, experimental (Fig.~\ref{fig:Coupled-maps-per}{\bf (a)}) and numerical (Fig.~\ref{fig:Coupled-maps-per}{\bf (b)}), 
show a remarkable concordance, even though we are using an outstanding resolution for both axis ($10^{-3}$). We highlight the
richness of the several coexisting motions in Fig.~\ref{fig:Coupled-maps-per} that are obtained from this method. The different 
dynamical behaviors are revealed explicitly in these diagrams, in contrast with the unresolved cases when using other indicators, 
as the Lyapunov exponents.

\begin{figure*}[htbp]
 \begin{center}
  \includegraphics[width=0.95\columnwidth]{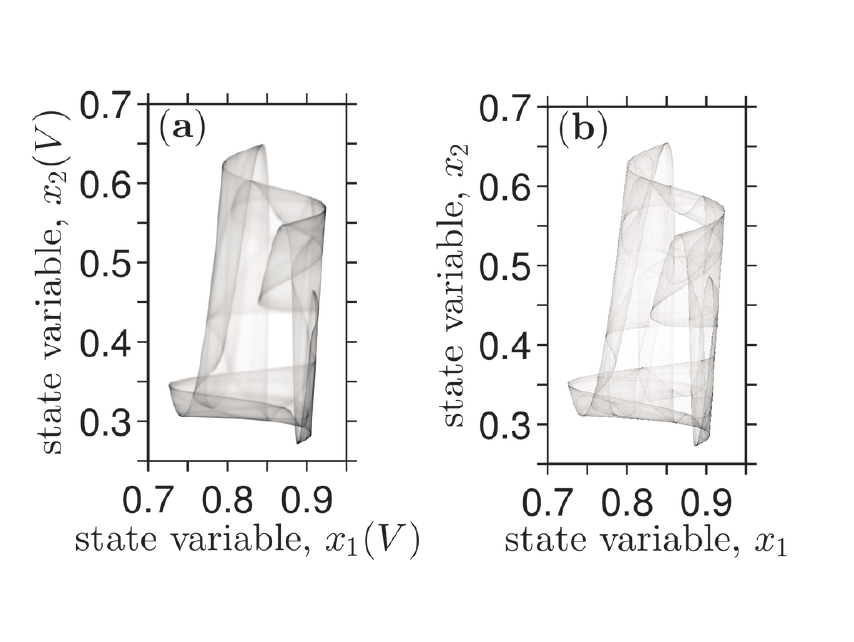} \hspace{1pc}
  \includegraphics[width=0.95\columnwidth]{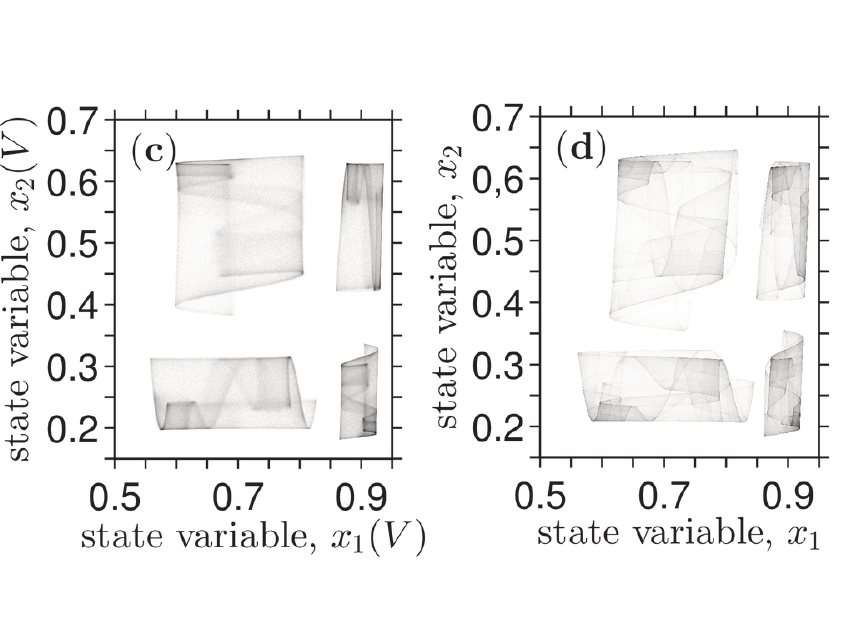}
 \end{center} \vspace{-2pc}
 \caption{ Phase space comparison between coupled logistic maps. Panels {\bf (a)} and {\bf (b)} show the attractors that are 
 found by fixing $r_1 = 3.22$, $r_2 = 3.79$, and $\epsilon = 0.937$, on the experimental and on the numerical coupled system, 
 respectively. Similarly, panels {\bf (c)} and {\bf (d)} show the experimental and numerical attractor for 
 $r_1 = 2.60$, $r_2 = 3.79$, and $\epsilon= 0.963$. The experimental time-series length to construct these portraits is of 
 $\sim 50 \times 10^6$, with an even larger numerical time-series length.}
 \label{fig:phase_space}
\end{figure*}

As a qualitative comparison, we show in Fig.~\ref{fig:phase_space} two phase-space portraits for the strongly
coupled system ($\epsilon \sim 1$) on two particular scenarios. The resulting attractors show remarkable complexity, 
nevertheless, the experimental (panels {\bf (a)} and {\bf (c)}) and numerical (panels {\bf (b)} and {\bf (d)}) attractors 
still remain remarkably close. This conclusion holds for a wide range of parameters and other phase-space portraits, which
we are omitted here.

%%%%%%%%%%%%%%%%%%%%%%%%%%%%%%%%%%%%%%%%%%%%%%%%%%%%%%%%%%%%%%%%%%%%%%%%%%%%%%%%%%%
\section{Conclusion}
\label{sec:con}
In this paper, we showed that our simple implementation represents correctly the behavior of the logistic map, and brings 
more possibilities for the study of chaos dynamics than previous implementations. We also implemented a Kaneko coupling, 
which we use to show that our logistic-map design allows us to couple several maps. In particular, we made a thorough analysis
of the coupled dynamics of two logistic-maps by critically comparing numerical and experimental data using Lyapunov exponents, 
orbit's periodicity, and phase-space portraits. Instead of logistic maps, different functions could be equally implemented using 
our approach. 

We can easily expand the system by adding extra couplings and maps since the coupling between two maps performed faultlessly. 
In particular, when studying the dynamics of two coupled maps, we observed that chaotic synchronization was possible in a wide 
range of coupling strengths, thus showing the robustness of the synchronous behavior and opening the question of how will a 
large set of coupled oscillators behave in an experimental system. The increase in size of our electronic design by the inclusion 
of more maps would allow us to make a full working network and study its behaviors, which are widely studied theoretically and
numerically but generally lack experimental study. In a field dominated by numerical simulations, our electronic design allows us 
to have a flexible system for future studies. Also, the implementation of a working network of many interacting chaotic oscillators 
could result in a system (if well calibrated) presenting high-dimensional chaotic dynamics, useful in a wide variety of applications
\cite{kaneko1.4916925}.

%%%%%%%%%%%%%%%%%%%%%%%%%%%%%%%%%%%%%%%%%%%%%%%%%%%%%%%%%%%%%%%%%%%%%%%%%%%%%%%%%%%

% 
% \begin{acknowledgements}
We acknowledge financial support from Programa de Desarrollo de las Ciencias B\'asicas (PEDECIBA), Uruguay. 
% \end{acknowledgements}


\begin{thebibliography}{10}
\expandafter\ifx\csname url\endcsname\relax
  \def\url#1{\texttt{#1}}\fi
\expandafter\ifx\csname urlprefix\endcsname\relax\def\urlprefix{URL }\fi
\expandafter\ifx\csname href\endcsname\relax
  \def\href#1#2{#2} \def\path#1{#1}\fi

\bibitem{Barabasi_2002}
A.-L. Barabasi, Linked: How everything is connected to everything else and what
  it means, Plume Editors, 2002.

\bibitem{Strogatz_2003}
S.~H. Storgatz, Sync: the emerging science of spontaneous order, Hyperion,
  2003.

\bibitem{Barrat_2008}
A.~Barrat, M.~Barthelemy, A.~Vespignani, Dynamical processes on complex
  networks, Cambridge University Press, 2008.

\bibitem{lorenz1962}
E.~N. Lorenz, Deterministic nonperiodic flow, J. Atmospheric Sci. 20 (1963)
  130--141.

\bibitem{winfree_2001}
A.~T. Winfree, The geometry of biological time, Vol.~12, Springer Science and
  Business Media, 2001.

\bibitem{kaneko1.4916925}
K.~Kaneko, From globally coupled maps to complex-systems biology, Chaos 25~(9).

\bibitem{Lloyd_1995}
A.~L. Lloyd, The coupled logistic map: A simple model for the effects of
  spatial heterogeneity on population dynamics, J. Theor. Biol. 173 (1995)
  217--230.

\bibitem{kendall1998spatial}
B.~E. Kendall, G.~A. Fox, Spatial structure, environmental heterogeneity, and
  population dynamics: analysis of the coupled logistic map, Theoretical
  population biology 54~(1) (1998) 11--37.

\bibitem{may1976simple}
R.~M. May, et~al., Simple mathematical models with very complicated dynamics,
  Nature 261~(5560) (1976) 459--467.

\bibitem{feigenbaum_1978}
M.~J. Feigenbaum, Quantitative universality for a class of nonlinear
  transformations, J. Stat. Phys. 19~(1).

\bibitem{Yorke1996}
K.~T. Alligood, T.~D. Sauer, J.~A. Yorke, Chaos: An Introduction to Dynamical
  Systems, Springer, 1996.

\bibitem{collet2009iterated}
P.~Collet, J.-P. Eckmann, Iterated maps on the interval as dynamical systems,
  Springer Science \& Business Media, 2009.

\bibitem{mcgonigal1987generation}
G.~McGonigal, M.~Elmasry, Generation of noise by electronic iteration of the
  logistic map, Circuits and Systems, IEEE Transactions on 34~(8) (1987)
  981--983.

\bibitem{phatak_1995}
S.~C. Phatak, S.~Suresh~Rao, Logistic map: A possible random number generator,
  Phys. Rev. E 51.

\bibitem{stone1993period}
L.~Stone, Period-doubling reversals and chaos in simple ecological models,
  Nature 365~(6447) (1993) 617--620.

\bibitem{pareek2006image}
N.~K. Pareek, V.~Patidar, K.~K. Sud, Image encryption using chaotic logistic
  map, Image and Vision Computing 24~(9) (2006) 926--934.

\bibitem{singh2010chaos}
N.~Singh, A.~Sinha, Chaos-based secure communication system using logistic map,
  Optics and Lasers in Engineering 48~(3) (2010) 398--404.

\bibitem{borujeni2015modified}
S.~E. Borujeni, M.~S. Ehsani, Modified logistic maps for cryptographic
  application, Applied Mathematics 6~(05).

\bibitem{baptista_1996}
M.~S. Baptista, I.~Caldas, Dynamics of the kicked logistic map, Chaos, Solitons
  and Fractals 7~(3).

\bibitem{baptista_1997}
M.~S. Baptista, I.~Caldas, The parameter space structure of the kicked logistic
  map and its stability, Int. J. Bif. Chaos 7~(2).

\bibitem{campos2011family}
E.~Campos-Cant{\'o}n, R.~Femat, A.~Pisarchik, A family of multimodal dynamic
  maps, Communications in Nonlinear Science and Numerical Simulation 16~(9)
  (2011) 3457--3462.

\bibitem{radwan2013some}
A.~G. Radwan, On some generalized discrete logistic maps, Journal of advanced
  research 4~(2) (2013) 163--171.

\bibitem{kaneko_1983}
K.~Kaneko, Transition from torus to chaos accompanied by frequency lockings
  with symmetry breaking, Progress of Theoretical Physics 69~(5).

\bibitem{kaneko1990clustering}
K.~Kaneko, Clustering, coding, switching, hierarchical ordering, and control in
  a network of chaotic elements, Physica D: Nonlinear Phenomena 41~(2) (1990)
  137--172.

\bibitem{maistrenko_1998}
Y.~L. Maistrenko, V.~L. Maistrenko, A.~Popovich, E.~Mosekilde, Transverse
  instability and riddled basins in a system of two coupled logistic maps,
  Phys. Rev. E 57~(3).

\bibitem{PhysRevE.92.052912}
L.~Q. English, Z.~Zeng, D.~Mertens,
  \href{http://link.aps.org/doi/10.1103/PhysRevE.92.052912}{Experimental study
  of synchronization of coupled electrical self-oscillators and comparison to
  the sakaguchi-kuramoto model}, Phys. Rev. E 92 (2015) 052912.
\newblock \href {http://dx.doi.org/10.1103/PhysRevE.92.052912}
  {\path{doi:10.1103/PhysRevE.92.052912}}.
\newline\urlprefix\url{http://link.aps.org/doi/10.1103/PhysRevE.92.052912}

\bibitem{horowitz1989art}
P.~Horowitz, W.~Hill, The art of electronics, Cambridge Univ. Press, 1989.

\bibitem{suneel2006electronic}
M.~Suneel, Electronic circuit realization of the logistic map, Sadhana 31~(1)
  (2006) 69--78.

\bibitem{garcia2013difference}
M.~Garc{\'\i}a-Mart{\'\i}nez, I.~Campos-Cant{\'o}n, E.~Campos-Cant{\'o}n,
  S.~{\v{C}}elikovsk{\`y}, Difference map and its electronic circuit
  realization, Nonlinear Dynamics 74~(3) (2013) 819--830.

\bibitem{cromer_1989}
S.~T. Welstead, T.~L. Cromer, Coloring periodicities of two-dimensional
  mappings, Comput. and Graphics.

\bibitem{amil2015electronic}
P.~Amil, C.~Cabeza, A.~C. Mart{\'i}, Exact discrete-time implementation of the
  mackey-glass delayed model, Circuits and Systems II: Express Briefs, IEEE
  Transactions on 62~(7) (2015) 681--685.

\bibitem{amil2015organization}
P.~Amil, C.~Cabeza, C.~Masoller, A.~C. Mart{\'i}, Organization and
  identification of solutions in the time-delayed mackey-glass model, Chaos: An
  Interdisciplinary Journal of Nonlinear Science 25~(4).

\end{thebibliography}
\end{document}